\documentclass[a4paper,11pt]{article}
\usepackage{jinstpub} 
\usepackage{lineno}
\usepackage{gensymb}
\usepackage{amsmath}
\usepackage{subcaption}
\usepackage{booktabs}
\usepackage[table]{xcolor} 
\usepackage{siunitx}
\usepackage{hyperref}
\definecolor{lightgray}{gray}{0.9}
\usepackage{tabularray}
\usepackage{todonotes}

\title{Characterization of the R14374-10 photomultiplier tube for the Pacific Ocean Neutrino Experiment}

\author[a]{O.~Aghaei,}
\author[b]{M.~Agostini,}
\author[b]{A.~Alexander Wight,}
\author[c]{K.~Ba\c{s},}
\author[a]{N.~Baily,}
\author[d,e]{P.~S.~Barbeau,}
\author[a]{A.~J.~Baron,}
\author[c]{S.~Bash,}
\author[c]{C.~Bellenghi,}
\author[c]{M.~Boehmer,}
% ...

\emailAdd{v.gousy-leblanc@tum.de}
\collaboration{The P-ONE collaboration}
\abstract{The Pacific Ocean Neutrino Experiment (P-ONE) is a planned cubic-kilometer-scale neutrino telescope that will be deployed in the Pacific Ocean, leveraging deep-sea infrastructure operated by Ocean Networks Canada (ONC). The first detector line will deploy 320 3-inch Hamamatsu R14374-10 photomultiplier tubes (PMTs) to detect Cherenkov light produced by charged particles propagating through water, as well as light from calibration sources and optical backgrounds. This article presents the systematic characterization of all PMTs to be integrated into the first detector line prior to deployment, ensuring compliance with all detector acceptance criteria. All PMTs satisfy the specified performance criteria: the average transit-time spread (TTS) was measured to be $1.73\pm 0.15\,$ns, and the average charge resolution at a gain of \num{5e6} was $0.43\pm 0.03\,$pC; the average after-pulse rate was measured to be around 6$\,\%$. The integrated P-ONE data acquisition (DAQ) system was also tested and cross-validated against a laboratory digitization system. The second part of the paper presents an in-depth characterization of a subset of six PMTs, focusing on the key properties required for detector simulations, including linearity, after-pulse rate, quantum efficiency, and relative collection efficiency. PMT linearity was measured at a gain of \num{5e6} using the complete P-ONE readout electronics, and the response was found to remain linear for signals up to 150 photoelectrons. After-pulse contributions from $\text{Cs}^{+}$, $\text{He}^{+}$, and other ions were identified in the subset. The peak quantum efficiency was measured to be $27.0\pm0.7\,\%$ at 395\,nm. Using measurements of the relative detection efficiency, the collection efficiency was found to differ by a wavelength-independent offset of 10--15$\%$ across the measured wavelength range.
}

\begin{document}
\maketitle
\flushbottom

\section{Introduction}
\begin{figure}[h]
    \centering
    \begin{subfigure}[b]{0.45\textwidth}
        \centering
        \includegraphics[width=\textwidth]{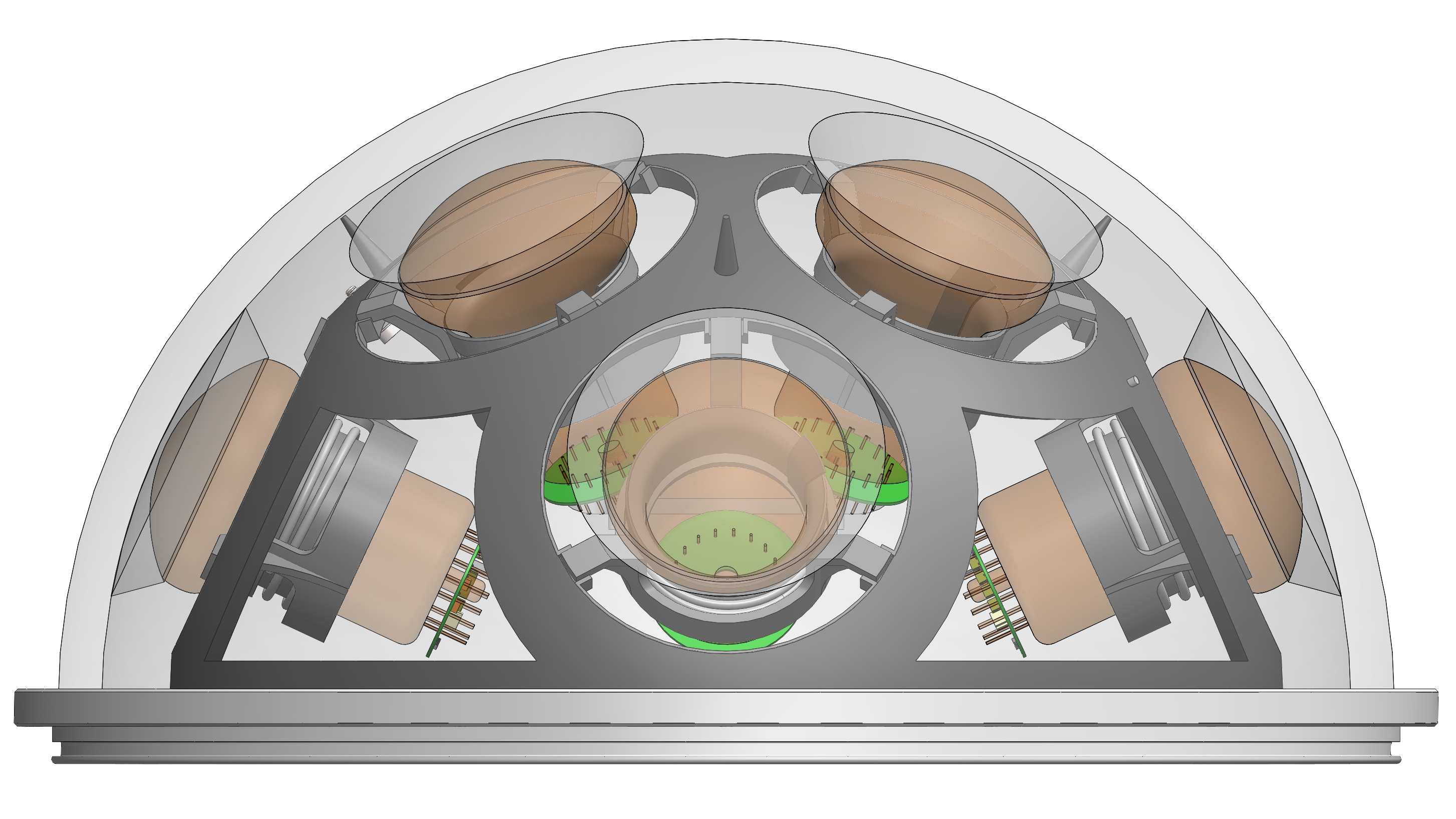}
        \caption{P-ONE optical module (P-OM)}
        \label{fig:sub1}
    \end{subfigure}
    \hfill
    \begin{subfigure}[b]{0.45\textwidth}
        \centering
        \includegraphics[width=\textwidth]{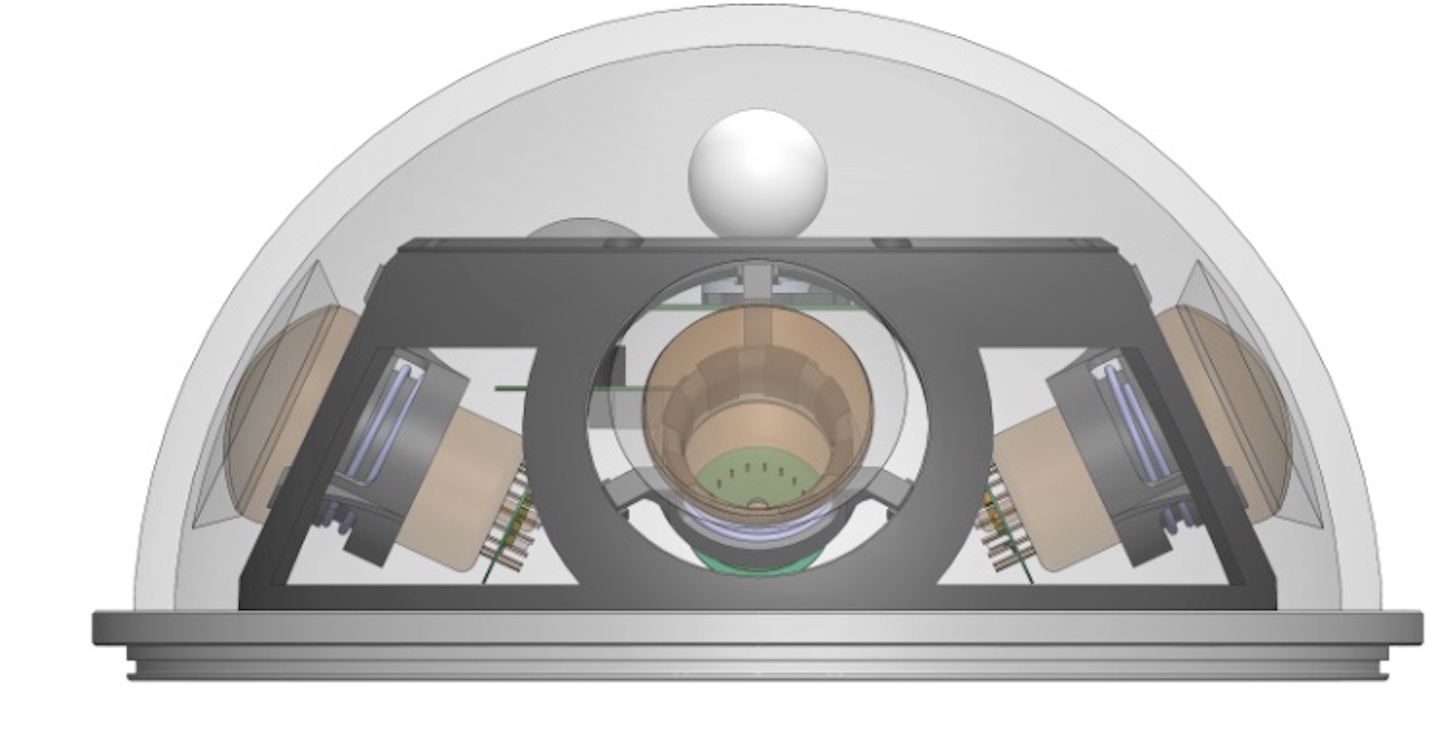}
        \caption{P-ONE calibration module (P-CAL)}
        \label{fig:sub2}
    \end{subfigure}
    \caption{The two optical modules for the first P-ONE detector line, a) shows a P-OM, which contains 16 PMTs (8 PMTs per hemisphere) and flasher inserts, while b) shows a P-CAL which contains an isotropic light source and eight PMTs (4 PMTs per hemisphere); taken from \cite{henningsenPacificOceanNeutrino2023}.}
    \label{fig:p_om_p_cal}
\end{figure}

The Pacific Ocean Neutrino Experiment (P-ONE) is a planned neutrino telescope in the Pacific Ocean. It will use the ocean water as a medium for detecting Cherenkov radiation from charged particles produced when high-energy neutrinos interact with it. Communication with the shore will be established through the NEPTUNE ocean observatory, a 850\,km-long loop of optical fiber network developed and operated by Ocean Networks Canada (ONC). The scientific objectives of the detector include searching for neutrino point sources, measuring the flavor-sensitive diffuse neutrino flux, and studying neutrinos originating from our galaxy \cite{Park2025Canadian,agostiniPacificOceanNeutrino2020b}. In addition, P-ONE offers the potential to support oceanographic measurements, such as pH and oxygen concentration, along the 1\,km vertical water column, providing a unique opportunity for long-term ocean monitoring and climate-related studies \cite{OceansProceeding}.
Before the deployment of the full array, the pathfinder experiments STRAW \cite{baillyTwoYearOpticalSite2021} and STRAW-b \cite{holzapfelBioluminescencePacificOcean} studied the feasibility of deploying a neutrino telescope in the Cascadia Basin, very close to the site at which the full P-ONE array is planned to be deployed. Following lessons learned from these pathfinder missions, the first P-ONE detector line was designed and constructed over the past few years and is currently ready for deployment. It comprises 18 optical modules (P-OMs), one calibration module (P-CAL) \cite{agostiniOpticalCalibrationSystems2026}, and one hybrid (P-OM/P-CAL) module. The P-OM and P-CAL hemisphere designs are shown in Figure \ref{fig:p_om_p_cal}. The P-OM consists of 16 Hamamatsu R14374-10 PMTs encased in a 17-inch glass housing made of 2 hemisphere. 

This article describes the characterization of photomultiplier tubes (PMT) for integration into P-ONE first line. First, Section \ref{sec:deep_sea} provides a brief overview of the deep-sea environment where the first P-ONE detection line will be deployed, along with a summary of relevant PMT parameters, followed by the acceptance criteria and selection. Section \ref{sec:testing_pmt} discusses the testing performed for all PMTs integrated into the modules of the first P-ONE line. Section \ref{sec:detailed} describes detailed measurements on a subset of PMTs using a picosecond laser and targeting photon-detection efficiency and quantum efficiency (Section \ref{sec:pde}), linearity (Section \ref{sec:linearity}), and after-pulsing (Section \ref{sec:afterpulse}). 

\section{Requirements and site conditions}
\label{sec:deep_sea}
The deep-sea environment at the P-ONE site directly shapes the performance requirements for the photomultiplier tubes and their readout. The Pacific Ocean is a dynamic medium for a neutrino telescope: currents move the detector lines, requiring continuous position monitoring \cite{collaborationPrototypeAcousticPositioning2025}; sediment and biological material can accumulate on the modules \cite{aghaeiLongTermStudy2025}; and microorganisms can emit light when disturbed, producing an important optical background (bioluminescence) \cite{liPolychromaticPopulationBioluminescent2026}. The STRAW \cite{baillyTwoYearOpticalSite2021} and STRAW-b \cite{holzapfelSTRAWbSTRingsAbsorption2024} pathfinder experiments provided crucial information for characterizing the two dominant sources of optical background in the Cascadia Basin: bioluminescence and light generated by radioactive decays of $^{40}\mathrm{K}$ in seawater.

\subsection{Deep-sea optical environment}
Seawater contains a uniform concentration of naturally occurring $^{40}\mathrm{K}$. Its decay produces electrons that can emit Cherenkov radiation in water, which are then detected by the PMT. The expected photon rate is directly correlated to the salinity of the water and is measured to be around 20\,kHz at the future P-ONE site \cite{baillyTwoYearOpticalSite2021}. This background can therefore be used to cross-calibrate the in situ detection efficiency of individual PMTs, as demonstrated by ANTARES \cite{melisInSituCalibrationKM3NeT2017}. The signature of $^{40}\mathrm{K}$ events is often identified by examining correlations among photosensors for low-intensity signals; see, e.g., \cite{melisInSituCalibrationKM3NeT2017}.  

Bioluminescence originates from marine organisms disturbed by turbulence generated by water flowing around the detector line. These organisms can emit light for 1 to 5 seconds across various wavelengths, including between 400/,nm and 500/,nm, where PMTs are sensitive. Bioluminescence can correlate with the water current and generate MHz rates of single-photon signals, as reported in \cite{gaertnerPacificOceanNeutrino2024,holzapfelBioluminescencePacificOcean}. Consequently, the P-ONE DAQ system must be capable of handling frequent PMT illumination caused by these optical backgrounds.

Beyond these optical backgrounds, the absorption and scattering properties of seawater also directly affect the propagation of Cherenkov light and, therefore, the detector performance. The long scattering length of ocean water -- see, e.g., \cite{baillyTwoYearOpticalSite2021} -- helps preserve a well-defined signature of particle tracks in the detector. Consequently, neutrino telescopes in water can achieve excellent angular resolution for point-source searches, provided that accurate synchronization between detector modules is ensured \cite{Twagirayezu2023PONEPerformance}. In this context, the timing resolution of the photosensors is a key factor in precisely reconstructing neutrino events from the secondary particles produced in the interaction.

\subsection{Photosensor requirements}
Knowing the different types of backgrounds, the characterization of the PMTs must cover not only their photon-detection capability and charge response but also their timing performance, response to high illumination rates, and the occurrence of delayed or spurious signals.

These requirements can be quantified through a set of PMT performance parameters. A detailed description of various PMT properties can be found in \cite{hamamatsu_handbook}, and the relevant parameters are briefly explained here. The photon detection efficiency (PDE) is the probability that an incident photon produces a detectable photoelectron, thereby characterizing the overall photon-detection capability of the PMT. The quantum efficiency (QE) is the probability that a photon incident on the photocathode generates a photoelectron, while the collection efficiency (CE) describes the probability that a photoelectron produced at the photocathode is successfully collected by the first dynode. Consequently, PDE depends on both QE and CE. The gain describes the amplification of the photoelectron signal through the dynode chain and determines the charge produced at the PMT output \cite{bellamyAbsoluteCalibrationMonitoring1994}. Its stability and charge resolution are therefore important for the detection and characterization of low-light signals. The timing performance is characterized by the transit time and its spread (TTS), with the latter describing the statistical variation in photoelectron arrival times at the PMT output, thereby contributing to the detector's timing resolution.

In addition to these primary response parameters, several effects can produce signals that are not directly associated with the prompt photoelectron signal. After-pulses are delayed signals that occur after the main pulse and are generally associated with processes such as ion feedback within the PMT, contributing to the detector background. Late pulses also occur after the main pulse and are generally associated with photoelectrons that follow trajectories resulting in longer transit times, such as backscattering from the first dynode. The dark current also contributes to the signal, since dark pulses are indistinguishable from single-photon signals and arise primarily from thermionic emission at the photocathode. Finally, linearity characterizes the relationship between the input light intensity and the measured PMT output signal. It is particularly important when the PMT is used to quantify detected light over a broad dynamic range, as deviations from linearity can introduce systematic effects in the reconstructed signal amplitude. In the present work, these parameters are experimentally characterized through dedicated measurements, providing a comprehensive description of the PMT response and its relevant performance characteristics for operation in the P-ONE detector.

\subsection{Photosensor selection criteria}
\label{sec:requirements}
Based on the environmental conditions and detector requirements discussed above, three main acceptance criteria were defined for PMT selection. First, the ambient optical background in the deep sea, dominated at low rates by $^{40}\mathrm{K}$ decays, allows for a relatively relaxed dark-rate requirement of $<3\,$kHz. Second, P-ONE's emphasis on pointing accuracy imposes a stringent requirement on timing performance, resulting in a TTS criterion of $<2.2\,$ns. Third, the after-pulse rate must remain below 10\% to limit delayed spurious signals that could contribute to the optical background. Among these parameters, TTS is the dominant timing property for PMT selection. A summary of the acceptance criteria is presented in Table \ref{tab:requirements}.
\begin{table}[h]
    \centering
    \begin{tabular}{lr}
        \toprule
        \textbf{Property} & \textbf{Acceptance crxiteria} \\
        \midrule
        Transit time spread (TTS) & $<2.2$\,ns \\
        Dark rate & $<3000$\,Hz \\
        After-pulse & $<10$\,\% \\
        \bottomrule
    \end{tabular}
    \caption{Acceptance criteria for the PMTs of the P-ONE line.}
    \label{tab:requirements}
\end{table}

Five different PMT models were initially investigated as candidates for the first P-ONE line \cite{gaertnerPacificOceanNeutrino2024,Elwinter}. For each model, the TTS and after-pulse rate were measured. The R14374 model was ultimately selected based on its low after-pulse rate, good timing performance, availability for large-scale procurement, and compatibility with the high-voltage base developed for IceCube \cite{andersonDesignPerformanceMultiPMT2021a}. Table \ref{tab:pmt_test} summarizes a selection of key parameters for the five candidate PMT models.

\begin{table}[h]
    \centering
    \begin{tblr}{
        colspec = {lcc},
        row{1} = {font=\bfseries},
        hline{1,Z} = {1pt},
    }
        PMT & TTS [ns] & After-pulse rate [\%] \\

        \textit{North Night Vision Technology (NNVT)} & & \\

        GDB-2031 & $2.0\pm0.1$ & 25 \\

        \textit{ET Enterprises} & & \\

        9323 KB & $3.6\pm0.2$ & 5 \\

        \textit{Hamamatsu} & & \\

        R15458-04 & $2.9\pm0.2$ & 8  \\
        \SetRow{bg=green!20}
        R14374    & $2.0\pm0.1$ & 8 \\
        R14689    & $2.2\pm0.1$ & 6 \\
    \end{tblr}
    \caption{Summary of the PMT models considered for the first P-ONE line, adapted from \cite{gaertnerPacificOceanNeutrino2024}. The dark-rate requirement of $<3\,$kHz was satisfied by all candidate models and was therefore not a discriminating criterion in the final PMT selection. The selected PMT is shown in green. }
    \label{tab:pmt_test}
\end{table}

\section{Production testing of PMTs for the first P-ONE line}
\label{sec:testing_pmt}
Following the selection of the PMT model, all sensors for the first P-ONE line were subjected to a common production and testing procedure. This section describes the PMT preparation, pre-integration testing, integration into the optical modules, and the subsequent post-integration characterization used to verify that PMT performance is preserved throughout the assembly process.

\subsection{PMT preparation and pre-integration testing}
\label{sec:module_assembly}
Prior to electrical testing and integration, each PMT from the production batch (320 units plus spares) was individually coated with an insulating film (\textit{Plastik 70}\footnote{\href{http://www.kontaktchemie.com/koc/KOCproductspecif.csp?CSPToken=Ju1fBUfWbVsHDweEtAiDJx$S6j8tPb5FeuVYxW1HUh_E0exGnhs0eKQ7qXJM4oDoiNbkp6exZe8s2WC$p7rnrQ--}{Plastik 70 -- Datasheet}}) applied to its neck. This coating is necessary because the PMTs operate at negative high voltage, which can cause electric charge to accumulate on the glass and lead to occasional light discharges, as reported by KM3NeT \cite{adrian-martinezMethodStabilisePerformance2016} and IceCube \cite{dittmerPerformanceStudiesNew2023}. Such discharges can increase the dark rate of individual PMTs and raise the background level within an optical module. The coating was applied by dipping each PMT into the insulating solution and leaving it to dry for 12 hours. The PMT output pins were subsequently cleaned to prevent contamination that could interfere with soldering the high-voltage base.

After applying the coating, the high-voltage base was integrated.
A version of the high-voltage base developed for the IceCube mDOM PMTs \cite{andersonDesignPerformanceMultiPMT2021a} was used for the first P-ONE line. Each base was soldered to the photosensor, and the assembly was tested before integration into an optical module. A PMT test facility developed for the IceCube Upgrade in Aachen \cite{IceCube:2024P07038,halveValidationOpticalSensors}, designed for the same 3-inch PMTs and high-voltage bases, was used to test all PMTs for the first P-ONE line individually. The measured PMT parameters are presented in Section \ref{sec:results} and referred to as the ``pre-integration measurement''.

\begin{figure}[h]
    \centering
    \includegraphics[width=1.0\linewidth]{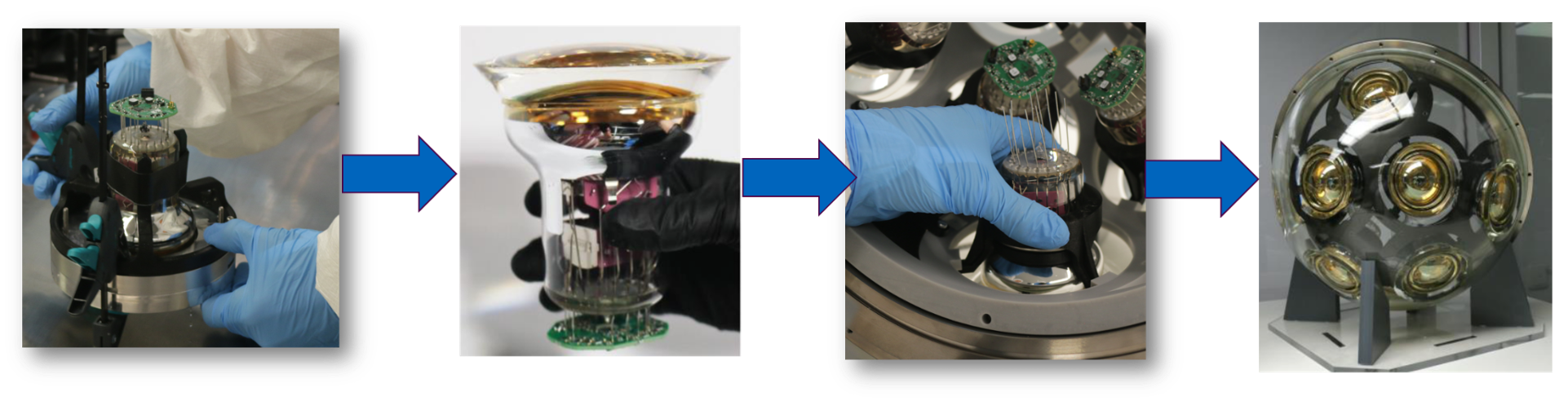}
    \caption{The first figure shows a PMT mounted in an aluminum mold designed to create a gel pad over the photocathode, as shown in the second image. The gel pad and PMT are then optically coupled to the inner surface of a glass hemisphere using an interface gel. Finally, the gel pad is secured within a 3D-printed matrix frame with a spring-loaded mechanism.}
    \label{fig:production_steps}
\end{figure}

Once a PMT passes this initial test, a gel pad made from \textit{Wacker SilGel 612}\footnote{\href{https://www.wacker.com/h/en-us/c/wacker-silgel-612-ab/p/000007546}{Wacker SilGel 612 -- Datasheet}} is coupled to the PMT; see also Figure \ref{fig:production_steps}. The PMT is then integrated into the optical module and positioned inside the glass pressure housing using an additional interface gel (\textit{Wacker Elastosil 601}\footnote{\href{https://www.wacker.com/h/de-ba/c/elastosil-rt-601-ab/p/000005741}{Wacker Elastosil 601 -- Datasheet}}). The gel pad serves as a mechanical and optical interface between the PMT and the optical module housing glass, enhancing light collection \cite{FinkC}. The complete procedure is shown in Figure \ref{fig:production_steps}. Internal electronics and cabling are then installed to control and readout each PMT. 

\subsection{Post-integration PMT testing}
\label{sec:setup_post}
The goal of the testing is to verify the functionality of all PMTs after integration in the P-OM/P-CAL, build a library of high-resolution waveforms from the different PMTs, and quantify PMT properties at subsequent stages of integration.

\begin{figure}[h]
    \centering
    \includegraphics[width=0.7\linewidth]{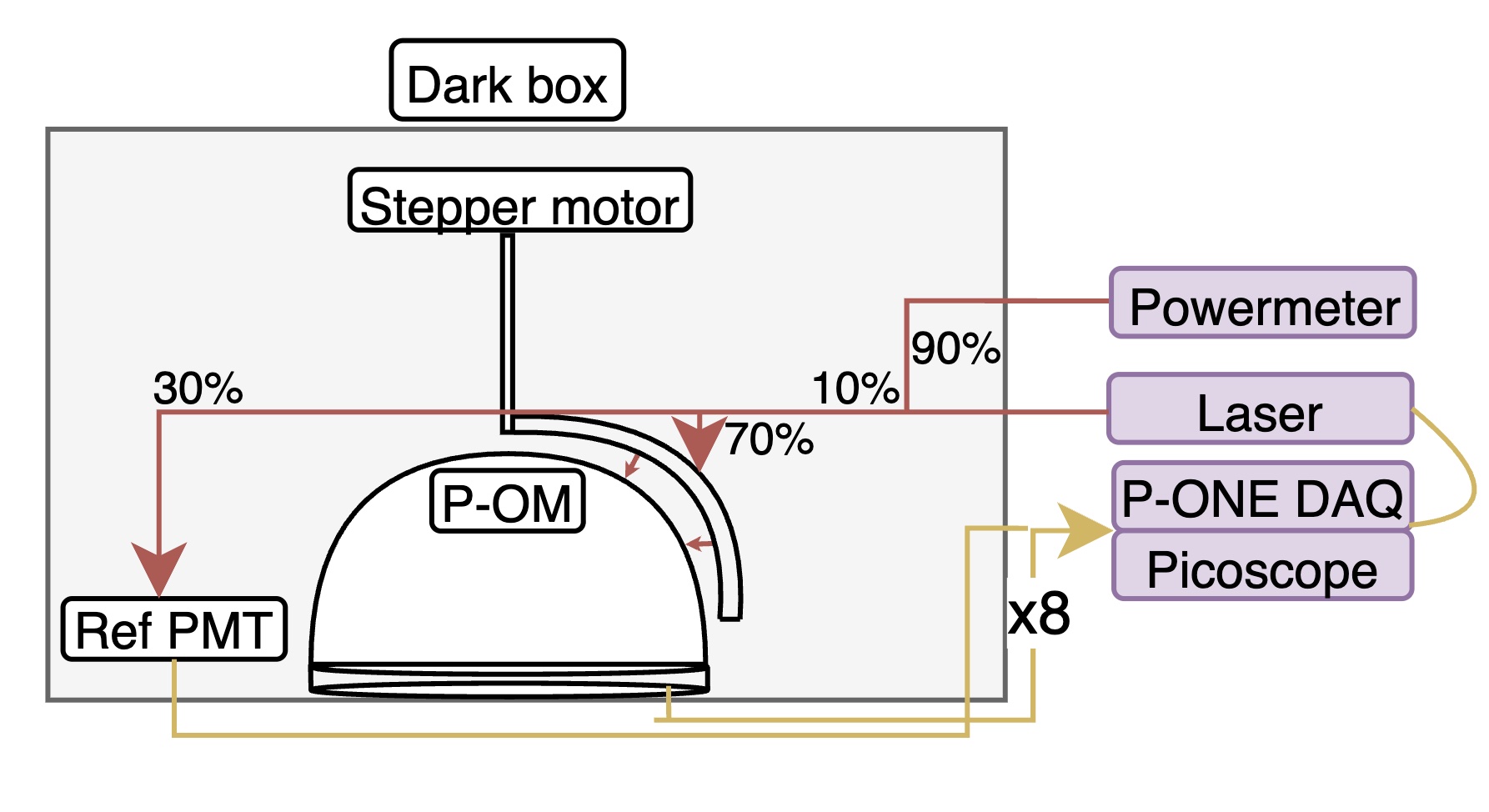}
    \caption{Left: schematic of the optical module test setup. Light is generated by a laser and subsequently split between a power meter, a reference PMT, and two PMTs housed within the optical module. Each PMT can be selected individually by rotating the stepper motor by 45°. Right: the actual setup, with the key components labeled; originally taken from \cite{gousy-leblancDevelopmentTestingOptical}.}
    \label{fig:p_om_testing_schematic}
\end{figure}

The schematic of the measurement setup is shown in Figure \ref{fig:p_om_testing_schematic}. It consists of a 405\,nm single-photon laser (\textit{PiLas}\footnote{\href{https://lf.hamamatsu.com/products/ultrafast-lasers/picosecond/pulsed-diode-lasers/pilas/}{PiLas -- Datasheet}}) connected to a 10/90 optical-fiber splitter for monitoring the laser output with a power meter (\textit{Newport 2936R}\footnote{\href{https://api.p1.mks.com/medias/sys_master/images/images/h3b/hb5/8796989751326/1936-R-2936-R-Power-and-Energy-Meter-Datasheet.pdf}{Newport 2936R -- Datasheet}}). The fiber is then split again 50/50 by another fiber splitter. Both fiber ends are then fixed to a 3D-printed mold connected to a stepper motor. The mold is aligned so that each fiber illuminates the center of a PMT when the motor is positioned at the correct angle. The PMT signals are then recorded by two ADCs: the internal P-ONE DAQ electronics, operating at approximately 210 MHz, and a high-precision ADC, Picoscope (\textit{PicoScope 6406E}\footnote{\href{https://www.picotech.com/download/datasheets/picoscope-6000e-series-data-sheet.pdf}{PicoScope 6406E -- Datasheet}}) operating at 1.25\,GHz.

\begin{figure}[h]
    \centering
    \includegraphics[width=0.8\linewidth]{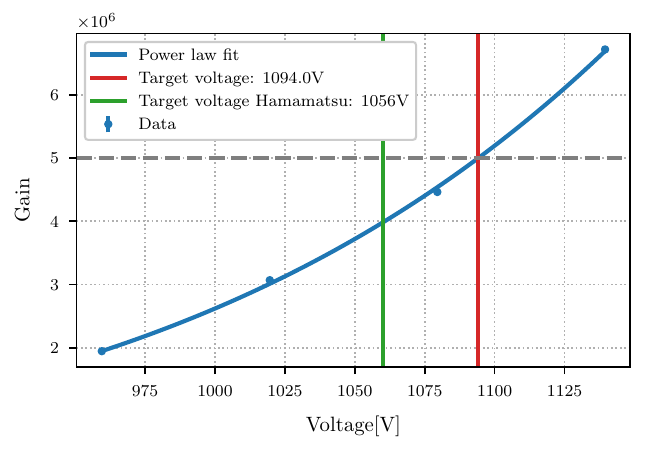}
    \caption{The gain of a PMT was measured for different voltages and then fitted with Equation \ref{eq:gain_calibration}. The reference voltage for a gain of \num{5e6} is found and defined as the target voltage. The original target voltage from Hamamatsu is also listed in green. }
    \label{fig:gain_calibration}
\end{figure}

\begin{figure}
    \centering
    \includegraphics[width=0.8\linewidth]{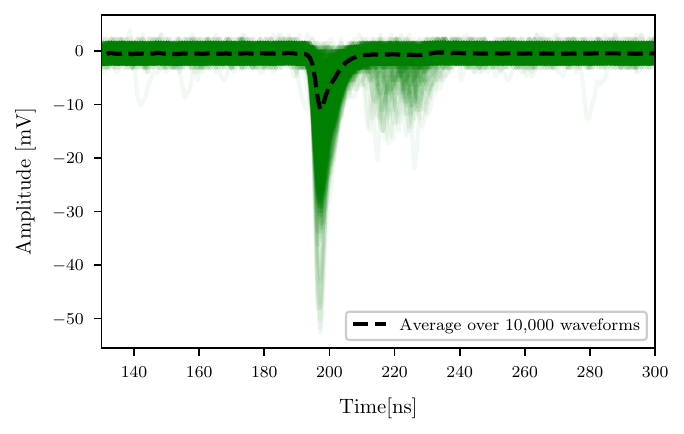}
    \caption{A collection of 10,000 waveforms recorded with the ADC is shown in green; the black line shows the average waveform. Some late pulses can be identified to the right of the main pulse, as well as some random light pulses, which are most likely dark counts or light background.}
    \label{fig:waveform_pico}
\end{figure}

The integrated hemisphere is inserted into the dark box using a platform and a movable crane.  
Once the laser is aligned with the PMT, an initial voltage is set to the manufacturer-recommended value for a gain of \num{5e6}. The gain is calibrated by selecting four bias voltages around the manufacturer's recommended nominal value and, for each voltage, measuring the gain as described in \cite{bellamyAbsoluteCalibrationMonitoring1994}. The gain dependence on voltage is fitted by the following Equation \cite{hamamatsu_handbook}:
\begin{equation}
\label{eq:gain_calibration}
    \text{G}(\text{V})=\text{a} \cdot \text{V}^\text{b},
\end{equation}
where V is the voltage, G is the gain, a is a scaling parameter, and b is a constant depending on the number of dynodes and the material used. The voltage is then extracted for a gain of \num{5e6}, which is P-ONE's target gain. Figure \ref{fig:gain_calibration} shows the result of such a fit, which includes the nominal values from the manufacturer. This is performed for both the higher- and lower-resolution digitizers. 
The timing distribution is constructed by fitting individual waveforms with a Gumbel distribution and extracting their mean. The FWHM of this distribution is then defined as the TTS. An example of the waveform obtained during the calibration can be seen in Figure \ref{fig:waveform_pico}.

\subsection{Post-integration validation and bulk results}
\label{sec:results}
One of the first validation steps is to verify that the PMT characteristics are not affected by integration into the optical module. This section compares selected PMT properties measured before and after integration and evaluates the performance of the P-ONE DAQ electronics \cite{daq_pone} against the Picoscope reference.

Pre-integration measurements were carried out at RWTH Aachen University at 3$\degree$C, as described in Section \ref{sec:module_assembly}. Post-integration measurements were performed with the setup described in Section \ref{sec:setup_post}, using both the Picoscope and the P-ONE DAQ electronics. In each summary plot, the bulk results are presented as differences relative to the post-integration Picoscope measurement.

\begin{figure}[h]
\centering
\includegraphics[width=1.0\linewidth]{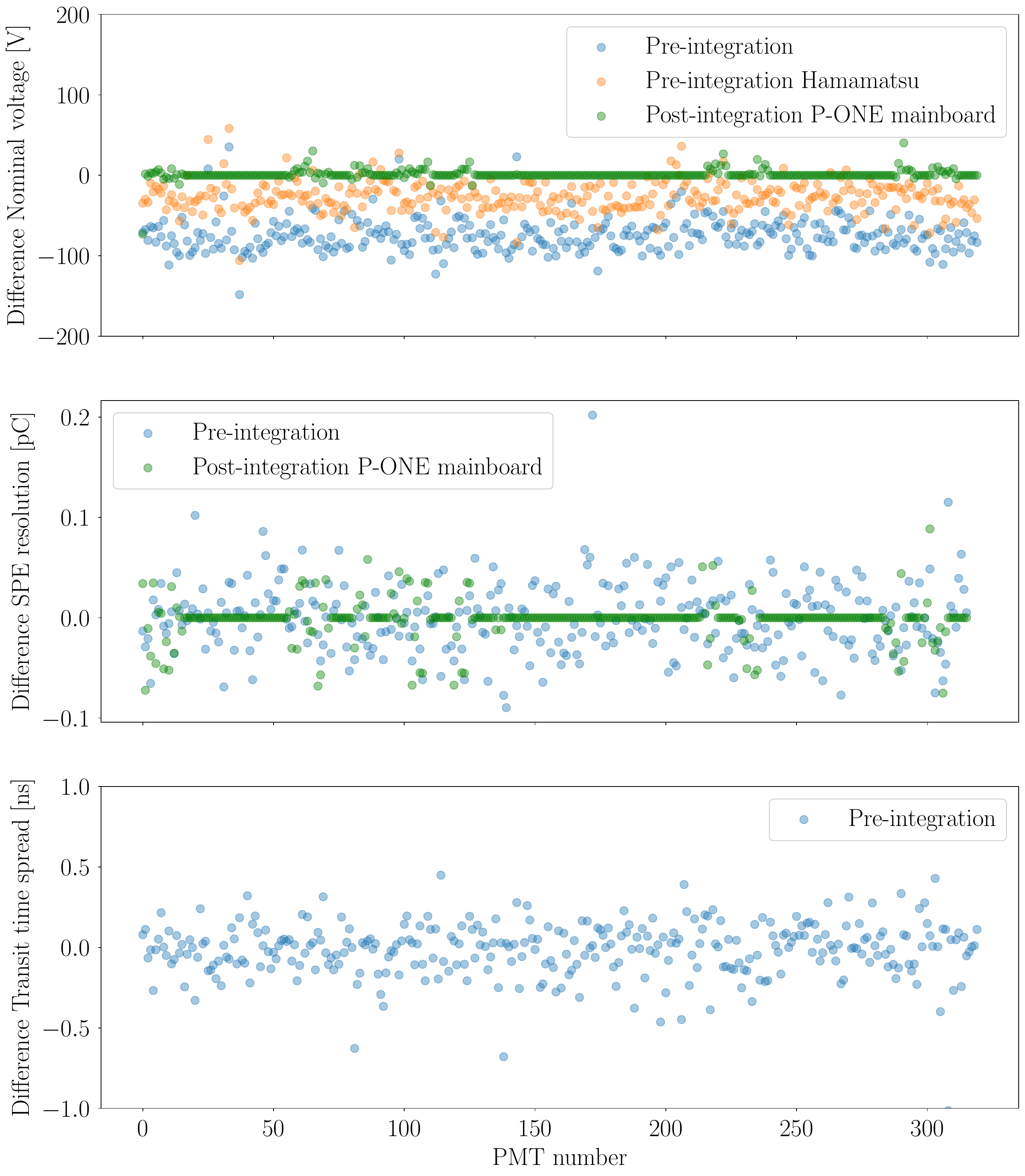}
\caption{Summary plot for all the PMTs tested, each data point corresponds to an individual PMT, identified by its serial number. Top: nominal voltage difference between the post-integration Picoscope measurements and both the pre-integration measurements and the P-ONE DAQ measurements. This voltage corresponds to a gain of \num{5e6}.
Middle: Difference in SPE resolution between the post-integration Picoscope measurements and both the pre-integration measurements and the P-ONE DAQ measurements.
Bottom: Difference in the TTS between the pre- and post-integration.} \label{fig:summary_large}
\end{figure}

The top panel of Figure \ref{fig:summary_large} shows the voltage difference required to reach a gain of \num{5e6} across the different testing phases. A small but consistent offset is observed between the pre-integration measurements and all subsequent datasets. This is likely due to the difference in measurement temperature: the post-integration measurements were performed at room temperature, whereas the pre-integration measurements were taken at \num{3}\degree C. PMT gain is known to be temperature-dependent \cite{lehmann2020pmt}. The P-ONE DAQ electronics show close agreement with the Picoscope measurements, confirming that the P-ONE DAQ can reliably reproduce these results. A smaller offset is also observed relative to the Hamamatsu reference data, which may be attributed to a residual temperature difference or differences in the measurement method. Importantly, all observed offsets can be compensated for by adjusting the high voltage applied to the PMT and therefore are all within the requirements.
The middle panel of Figure \ref{fig:summary_large} shows the difference in SPE resolution at a gain of \num{5e6}; good agreement is observed across all integration phases. This agreement indicates that the SPE response is preserved through integration and consistently reconstructed by both readout systems.
 The bottom panel of Figure \ref{fig:summary_large} shows the TTS measurements for all PMTs using both digitizers. These measurements are performed without correction for light-source jitter or contributions from the data acquisition system. Consequently, the reported values include contributions beyond the intrinsic PMT timing resolution. Fluctuations are observed between measurements, likely due to differences between the digitizers and the corresponding timing reconstruction.
 %
 Both sets of measurements are consistent within the limits defined in Section \ref{sec:requirements}; the final timing resolution will be determined by calibration with dedicated light sources integrated into each hemisphere. The mean values of the two main measured parameters are summarized in Table \ref{tab:avg_parameter}.
\begin{table}[h]
    \centering
    \begin{tabular}{lc}
        \toprule
        Parameter & Measurements \\
        \midrule
        Charge resolution [pC] & $0.42\pm0.03$ \\
        TTS [ns] & $1.73\pm0.15$ \\
        \bottomrule
    \end{tabular}
    \caption{Mean and spread of the parameters for all the tested PMTs using the Picoscope.}
    \label{tab:avg_parameter}
\end{table}

No PMTs were damaged during integration. The PMTs were randomly assigned during module assembly, avoiding selection bias and resulting in hemispheres with a representative distribution of TTS values. Finally, the P-ONE DAQ electronics were shown to provide charge-resolution and nominal-voltage measurements at a gain of \num{5e6} that are comparable to those obtained with the Picoscope. These results confirm that the integration procedure preserves PMT performance.

\section{Detailed characterization of selected PMTs}
\label{sec:detailed}
A detailed characterization was performed for a subset of PMTs. This included the measurement of quantum, photon-detection, and collection efficiencies; a linearity measurement; and after-pulsing. 

\subsection{Quantum, photon-detection, and collection efficiency}
\label{sec:pde}
 
\begin{figure}
    \centering
    \includegraphics[width=0.6\linewidth]{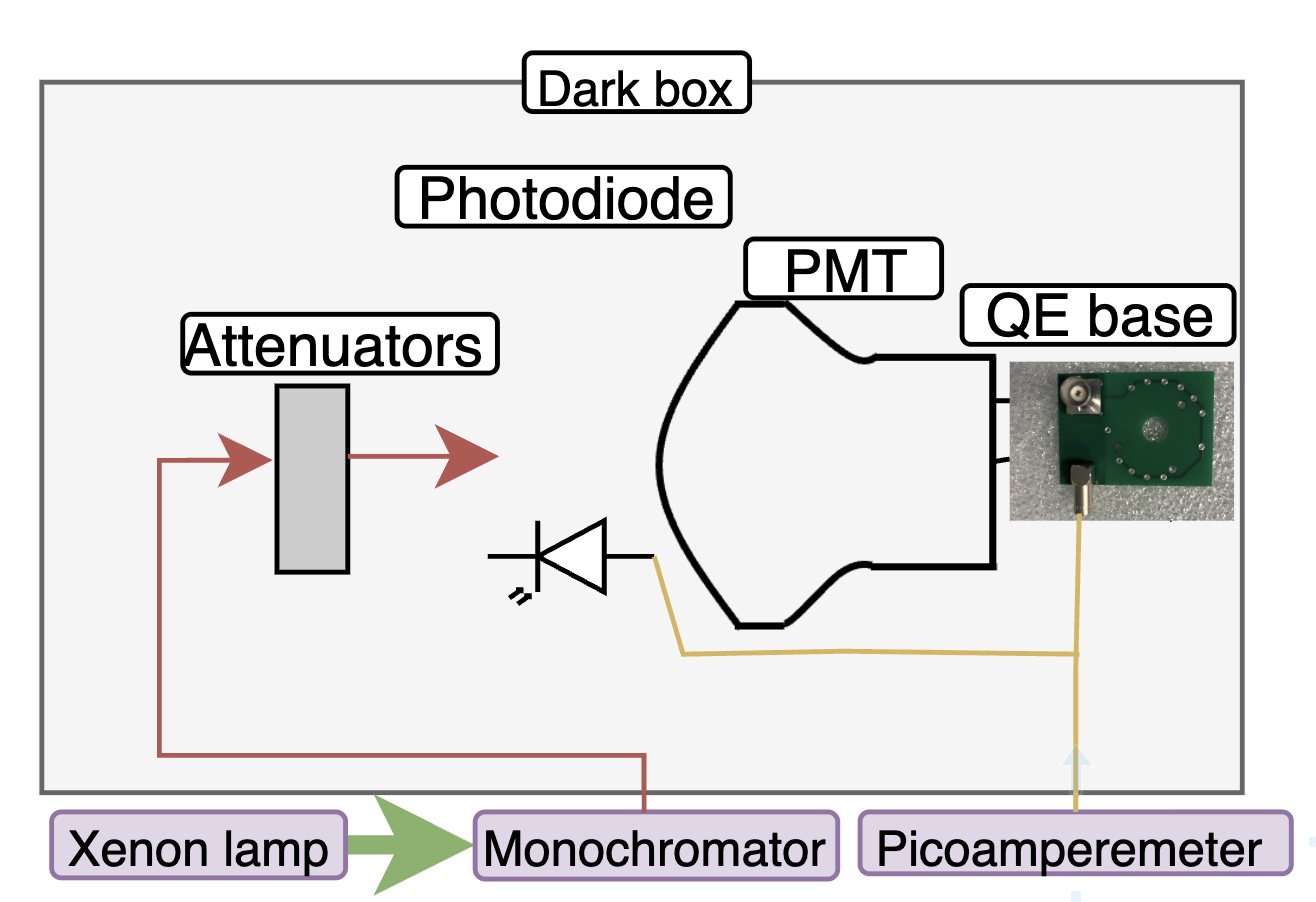}
    \caption{Schematic of the quantum efficiency measurement. A Xenon lamp is fiber coupled to a monochromator to select a wavelength, which is then attenuated via Neutral density (ND) filters before reaching the reference photodiode and PMT.}
    \label{fig:qe_setup}
\end{figure}

The QE is a fundamental property of the PMT. The schematic of the QE measurement setup is shown in Figure \ref{fig:qe_setup}.  A fiber-coupled Xenon lamp (\textit{Thorlabs SLS205}\footnote{\href{https://media.thorlabs.com/globalassets/items/s/sl/sls/sls205/ctn014628-d02.pdf?v=0116025423}{Thorlabs SLS205 -- Datasheet}}) is connected to a monochromator (\textit{Light source tech Monochromator-F}\footnote{\href{https://www.lightsource.tech/en/tunable-light-sources/the-monochromator-f/}{Light source tech Monochromator-F -- Datasheet}}), providing a stable light source over the wavelength range of 300–700\,nm. The monochromator output is passed through two 0.4\,mm fixed attenuators (\textit{Thorlabs FA58M}\footnote{\href{https://www.thorlabs.com/item/FA58M}{Thorlabs FA58M -- Datasheet}}) and a 3OD neutral density filter (\textit{Edmund optics 47-535}\footnote{\href{https://www.edmundoptics.de/p/30-od-25mm-dia-nir-nd-filter-/7616/}{Edmund optics 47-535r -- Datasheet}}), resulting in a total attenuation of \num{1e6}. A special PMT voltage base, powered by an external voltage supply, connects all the dynodes in series. The voltage is set to the expected voltage between the photocathode and the first dynode (250\,V). \\

 To minimize handling, the same light path is used for both the PMT response measurement and the calibration with the external reference — a NIST-calibrated photodiode (\textit{Hamamatsu S2281}\footnote{\href{https://www.hamamatsu.com/eu/en/product/optical-sensors/photodiodes/si-photodiodes/S2281.html}{Hamamatsu S2281 -- Datasheet}}) that provides an absolute photon flux \cite{NIST_Photodiode_Calibration}. This approach is viable because the xenon lamp spectrum is stable ($\leq$5$\%$ over 3h). The observed stability confirms that the QE measurement can be reliably performed .\\

Additionally, the currents from both the photodiode and the PMT are measured using the same picoamperemeter (\textit{Keithley 6485}\footnote{\href{https://www.tek.com/en/low-level-sensitive-and-specialty-instruments/series-6400-picoammeters-manual/model-6485-model-6487}{Keithley 6485 -- Datasheet}}) to reduce systematic uncertainty, and the dark current is subtracted from all measurements. To measure the QE, the absolute photon flux must be known. In the photodiode, it can be defined as, 
\begin{equation}
\label{eq:pd_eq_1}
    \text{N}_{\text{photon}}/\text{sec}=\frac{(I_{\mathrm{PD}} -I_{\mathrm{DKPD}})\cdot \lambda}{h\cdot c\cdot R(\lambda)},
\end{equation}
where $I_{\mathrm{PD}}$ is the measured current of the photodiode (A), $I_{\mathrm{DKPD}}$ the dark current measured when no light is emitted (A),  $\lambda$ is the wavelength (m), h is the Planck constant, c the speed of light and R($\lambda$) is the responsivity of the photodiode (A/W). 
Similarly, for the PMT, the number of photons detected is defined as,
\begin{equation}
\label{eq:photon_pmt}
    \text{N}_{\text{photon}}/\text{sec}=\frac{I_{\text{PMT}}-I_{\text{DKPMT}}}{\text{PDE} \cdot \text{G} \cdot \text{e}},
\end{equation}
where $I_{\text{PMT}}$ is the current measured on the PMT (A),$I_{\text{DKPMT}}$ is the dark rate of the PMT (A), G is the gain of the PMT, e is the electron charge and PDE is the photon detection efficiency. Using eq \ref{eq:pd_eq_1} and eq \ref{eq:photon_pmt}, PDE is then defined as,
\begin{equation}
\label{eq:pde}
\operatorname{PDE}(\lambda)=\frac{I_{\mathrm{PMT}} -I_{\mathrm{DKPMT}}}{I_{\mathrm{PD}}-I_{\mathrm{DKPD}}} \cdot \frac{R(\lambda)\cdot h\cdot c}{e \cdot \lambda \cdot  G},
\end{equation}
In this case, because a special base is used, the gain and collection efficiency are set to 1 since the dynodes are all connected,  hence obtaining
\begin{equation}
\label{eq:qe}
\operatorname{QE}(\lambda)=\frac{I_{\mathrm{PMT}} -I_{\mathrm{DKPMT}}}{I_{\mathrm{PD}}-I_{\mathrm{DKPD}}} \cdot \frac{R(\lambda)\cdot h\cdot c}{e \cdot \lambda }
\end{equation}

\begin{figure}[h]
    \centering
    \includegraphics[width=1.0\linewidth]{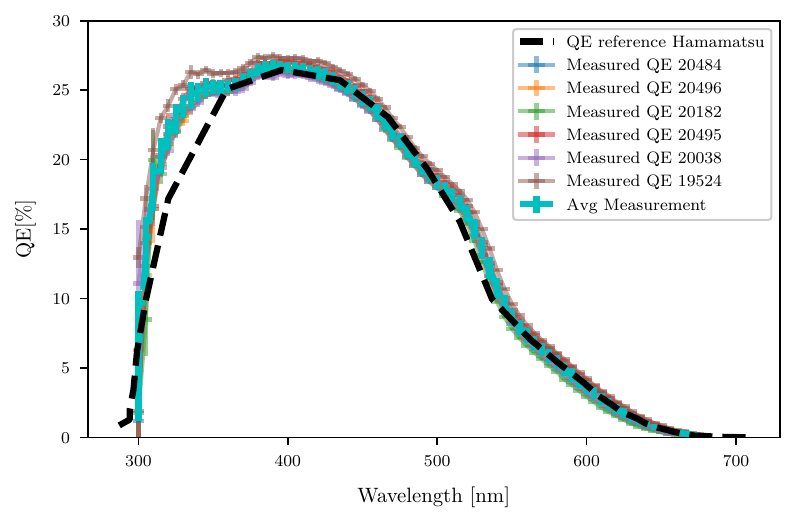}
    \caption{QE measurements of six PMTs, the average of all the efficiencies is shown in blue, the reference measurement Hamamatsu datasheet is also shown in black. }
    \label{fig:qe_measurement}
\end{figure}
\paragraph*{Results} The quantum efficiency of six PMTs is measured and shown with its uncertainty in Figure \ref{fig:qe_measurement}.  Figure \ref{fig:qe_measurement} also shows the mean QE measured across all tested PMTs.\\

Two features appear to deviate from the Hamamatsu reference measurements: a relative excess at 350\,nm, consistently observed across all tested PMTs, and a secondary increase in efficiency around 550\,nm. The first deviation occurs in a wavelength region characterized by larger measurement uncertainties and reduced sensitivity of the reference device. Both features may also be attributable to the limited sample size of six PMTs investigated in this study, as all units originated from the same production batch. In contrast, measurements reported by KM3NeT in \cite{collaborationEvaluationUpgraded3inch2025}, based on more than 200 PMTs, exhibit a substantial spread in measured efficiencies that encompasses both observed excesses. The highest quantum efficiency was measured at 395\,nm for $27.0\pm 0.7\,\%$ with a spread of 0.4$\%$ in the sample used. These measurements are in line with the expectations. \\

\begin{figure}[h]
\centering
\includegraphics[width=0.8\linewidth]{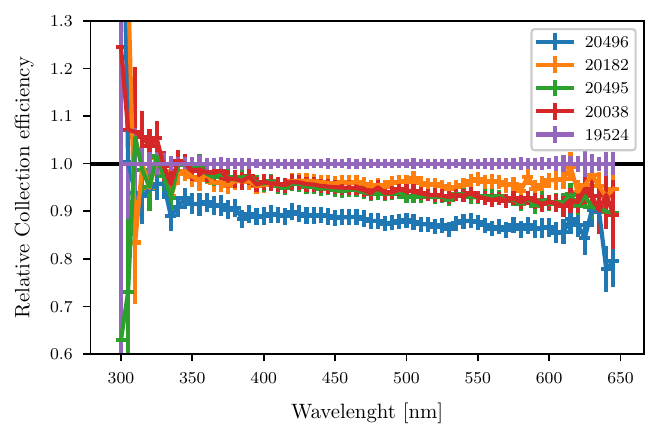}
\caption{Ratio of the collection efficiency measurement for six different PMTs. The reference PMT has a collection efficiency of 1 by definition, shown in purple. }
\label{fig:ce_ratio}
\end{figure}

In addition, a measurement of the photon-detection efficiency, or PDE, is particularly useful for detector simulations and for in-situ comparisons using $^{40}\mathrm{K}$ as a steady source of light \cite{melisInSituCalibrationKM3NeT2017}. The relative photon-detection efficiency was measured, and the relative collection efficiency was calculated from it. The setup is based on the one used for the QE measurement, with two main differences: the PMT high-voltage base is switched back to the one used in normal operation (full dynode chain) and is set to a gain of \num{2e6}. The intensity of the light is reduced by an additional neutral density filter  (\textit{Edmund optics 47-535}\footnote{\href{https://www.edmundoptics.de/p/30-od-25mm-dia-nir-nd-filter-/7616/}{Edmund optics 47-535r -- Datasheet}}).
The relative photon-detection efficiency is defined as: 
\begin{equation}
    \text{PDE}_{\text{relative}}=\frac{\text{PDE}}{\text{PDE}_{\text{ref}}}=\frac{(I_{\mathrm{PMT}} -I_{\text{DKPMT})} \cdot \text{G}_{\text{ref}} }{(I_{\mathrm{PMT \text{ref}}} -I_{\mathrm{DKPMT ref}})\cdot \text{G}}
    \label{eq:pde_rela}
\end{equation}
  The relative collection efficiency can then be calculated using equation \ref{eq:pde_rela} and the previously measured quantum efficiency giging :
\begin{equation}
    \frac{CE_{\text{PMT}}}{CE_{\text{ref}}}=\frac{\text{PDE}_{\text{relative}}}{\frac{\text{QE}}{\text{QE}_{\text{ref}}}}
\end{equation}
The expected collection efficiency for the PMT R14374 is above 85$\%$ for all wavelengths. The collection efficiency ratio is shown in Figure \ref{fig:ce_ratio} for the different PMTs, where all PMTs vary between 10 and 15$\%$ from the reference PMT. The error bars reflect error propagation in both the QE and PDE measurements, primarily due to the quantum efficiency measurement. Similar behavior has been observed by IceCube \cite{UnlandElorrieta2023}.

\subsection{Linearity}
\label{sec:linearity}

\begin{figure}[h]
\centering
\includegraphics[width=0.6\linewidth]{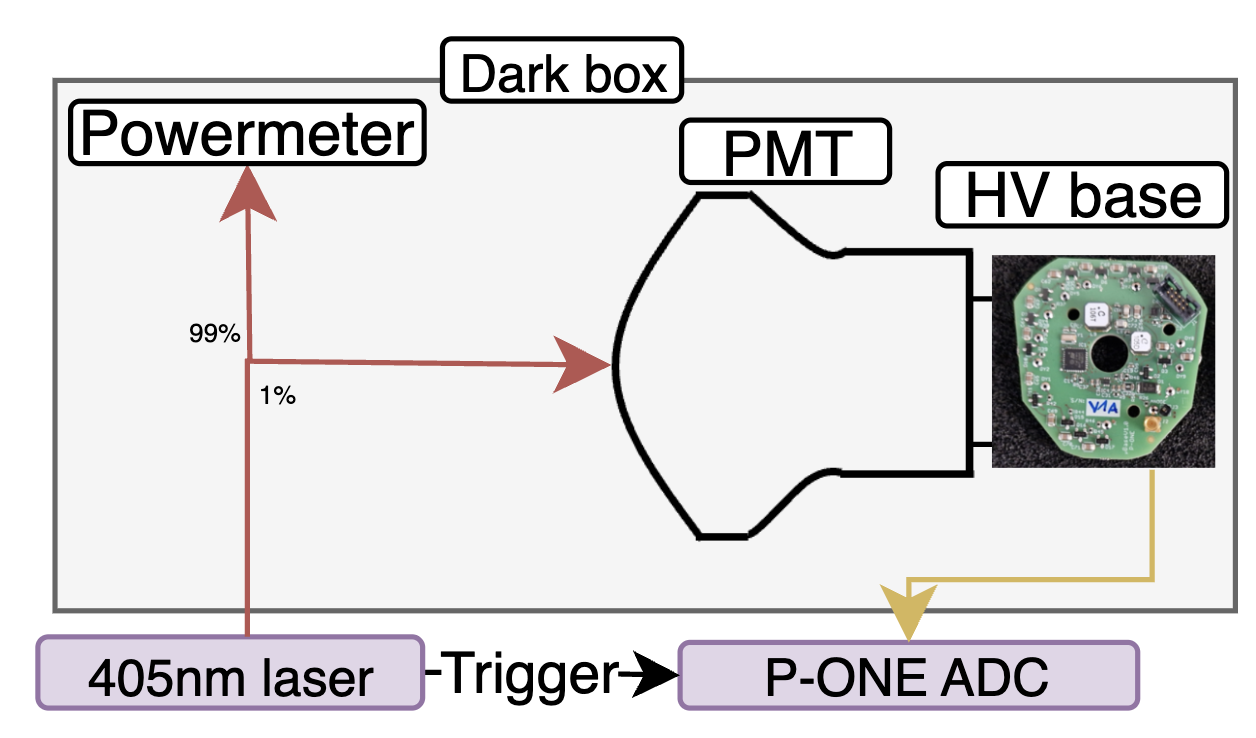}
\caption{Schematic of the linearity setup. The laser signal is split so that 99\% of the light is directed to the powermeter and 1$\%$ goes to the PMT.}
\label{fig:linearity_setup_schematic}
\end{figure}

The linearity of the PMT and P-ONE readout electronics must be characterized to determine the available photon dynamic range and quantify the impact of saturation on the measured signal. A fast-pulsed 405\,nm diode laser (\textit{PiLas}\footnote{\href{https://lf.hamamatsu.com/products/ultrafast-lasers/picosecond/pulsed-diode-lasers/pilas/}{PiLas -- Datasheet}}) is used to generate light pulses at a repetition rate of 10\,kHz. A fiber beam splitter directs 99$\%$ of the photons to a power meter (\textit{Newport 2936R}\footnote{\href{https://api.p1.mks.com/medias/sys_master/images/images/h3b/hb5/8796989751326/1936-R-2936-R-Power-and-Energy-Meter-Datasheet.pdf}{Newport 2936R -- Datasheet}}), while the remaining 1$\%$ are sent to the PMT connected to the P-ONE DAQ electronics. Each laser pulse produces a trigger signal, which is also fed to the P-ONE system, configured to acquire waveforms within a 160\,ns window following it. The light intensity is controlled by adjusting the laser drive current and is continuously monitored by the power meter. For each intensity setting, 1,000 waveforms are recorded. A schematic of the setup is shown in Figure \ref{fig:linearity_setup_schematic}.  

\begin{figure}[h]
    \centering
    \includegraphics[width=0.8\linewidth]{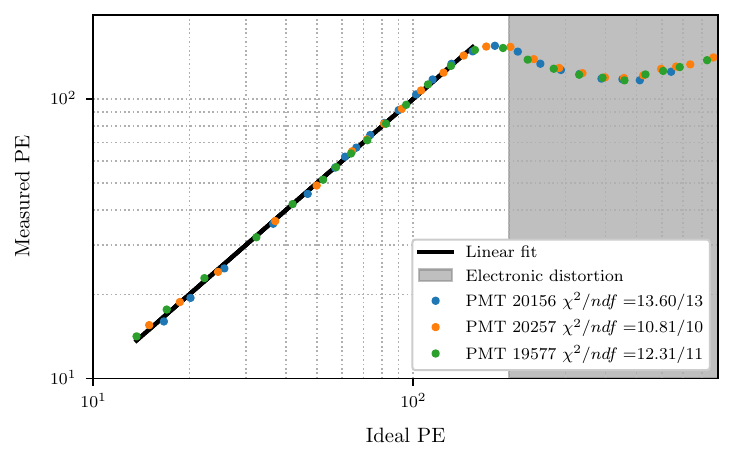}
    \caption{Linear fit for 3 different PMTs with the same channel in the P-ONE DAQ, with the number of photoelectrons (PE) for a gain of \num{5e6}. The 3 results are consistent and show that the PMT response is linear. The shaded region indicates saturation, which results in electronic ringing and pulse-shape distortion.}
    \label{fig:linearity_measurement_photons}
\end{figure}

For each light intensity, the number of detected photoelectrons (PE) is calculated. The photon flux on the PMT is calculated as,
\begin{equation}
\frac{\text{number of photoelectrons}}{\text{pulse}} = \frac{\text{integrated charge}}{\text{e} \cdot \text{G}},
\end{equation}
where the integrated charge is obtained from the digitized signal, (e) is the elementary charge, and (G) is the PMT gain. Once those quantities are calculated, the expected current and number of photoelectrons can be determined from the power meter readings, since the beam splitter ratio is known for each intensity. To measure the linearity of a PMT, the behavior between the calibrated photon flux and the measured number of PE is quantified using a linear fit. 
\paragraph*{Results}
 Figure \ref{fig:linearity_measurement_photons} shows the linearity measurement across 3 different PMTs at a gain of \num{5e6}. It can be seen that the PMTs behave identically in the linear regime up to about 150 photoelectrons, providing confidence that the PMT-to-PMT difference in linearity is small. The PMT response decreases after saturation, primarily due to electronic ringing at high charges, which results in pulse-shape distortions within the P-ONE DAQ electronics. 

\subsection{After-pulsing}
\label{sec:afterpulse}
The after-pulse rate is an important measurement for assessing the level of background originating within the PMT and 
identifying which residual gases may have entered the PMT during fabrication. Once inside the tube, these gases can be ionized by primary photoelectrons, releasing secondary electrons that generate delayed waveforms, defined as after-pulses \cite{hamamatsu_handbook}. Helium is a common contaminant because it can diffuse through the glass envelope and remain inside the PMT. Another frequently observed gas is hydrogen $\text{H}^+$, which can be introduced during PMT glass production. $\text{Cs}^+$ ions can come from the photocathode material itself as reported by the Hyper-Kamiokande collaboration \cite{wcte_pmt_presentation_tipp}. 
The type of residual gas present in the PMT can be identified by a characteristic second pulse with a particular delay time. This delay depends on the ion drift time, which can be calculated using the following equation \cite{maTimeAmplitudeAfterpulse2011} 
\begin{equation}
\label{eq:delay_after}
\text{t}=\frac{\pi}{4} \sqrt{\frac{2 \text{m}}{\text{e}V_0}}\text{L}
\end{equation}
where t is the delay time (sec), L is the distance between the photocathode and second dynode (56 mm for the current PMT ), $V_0$ is the electric potential (at the second dynode set to 450V), m is the mass (kg) of the ion, and e is the fundamental electron charge (Coulomb). 
\begin{table}[h]
    \centering
    \begin{tabular}{cc}
        \toprule
        \textbf{Gas} & \textbf{Delay time at second dynode [ns]} \\
        \midrule
        $\text{H}^+$ & 293 \\
        $\text{He}^+$ & 586 \\
        $\text{CH}_4$ & 1172 \\
        $\text{H}_2\text{O}_+$ & 1393 \\
        $\text{N}_2$ & 1551 \\
        $\text{Ar}^+$ & 1830 \\
        $\text{Cs}^+$ & 3379 \\
        \bottomrule
    \end{tabular}
    \caption{Predicted after-pulse time for the most likely gases  found in the PMT from Equation\,\ref{eq:delay_after}.}
    \label{tab:predicted time}
\end{table}

An optical fiber connected to the picosecond laser described in Section~\ref{sec:setup_post} was positioned at the center of the PMT. The laser trigger and the PMT output were both connected to a 1.25\,GHz Picoscope (\textit{PicoScope 6406E}\footnote{\href{https://www.picotech.com/download/datasheets/picoscope-6000e-series-data-sheet.pdf}{PicoScope 6406E -- Datasheet}}), which recorded waveforms over a 1\,ms window following each trigger; 100,000 waveforms were recorded. The main pulse window is defined as [-10,+50]\,ns of the bin with the most events, which corresponds to the main signal. Any signal observed in the remaining time window, with an amplitude exceeding a 5\,mV threshold, was identified as an after-pulse. The after-pulse rate is then calculated as the ratio of the number of after-pulses to the number of main pulses. In addition, the time and charge of each pulse are recorded to identify the potential gas inside the detector. 

\begin{figure}[h]
    \centering
    \includegraphics[width=0.8\linewidth]{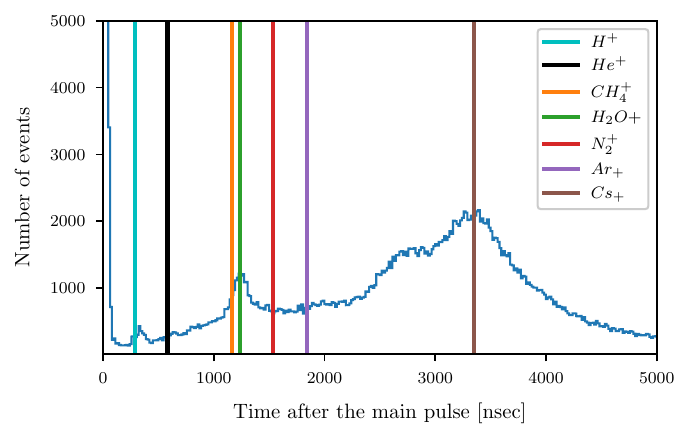}
    \caption{Timing spectrum of the after-pulse distribution for a single PMT. A threshold of 5mV is applied, and the time of each of those pulses is plotted in this histogram.}
    \label{fig:after_pulse_distribution}
\end{figure}

\paragraph*{Results}
The timing profile and charge deposition of after-pulses are shown in Figure \ref{fig:after_pulse_distribution} for a single PMT. The different gases predicted in Table \ref{tab:predicted time} are traced, and $\text{H}^+$, $\text{H}_2\text{O}_+$ and $\text{Cs}^+$ gas can be identified. This result is consistent with the Hyper-Kamiokande measurement from the same model \cite{wcte_pmt_presentation_tipp}. However, as described before, each gas authentication is subject to large uncertainty; therefore, when two peaks are quite close (e.g., for $\text{CH}_4$ and $\text{H}_2\text{O}_+$), it is difficult to confirm which gas is present. 
 The after-pulse rate was measured for 320 PMTs at RWTH Aachen before their integration into the optical module. The count of after-pulses and main pulses was calculated using the analysis in \cite{halveValidationOpticalSensors}, resulting in an average value of 6$\%$ with a spread of 4$\%$. Similar results are reported by KM3NeT in \cite{collaborationEvaluationUpgraded3inch2025}. 

\section{Summary and outlook}
In this paper, we presented measurements of the R14374-10 photomultiplier tubes (PMTs) used in the first P-ONE detector line, characterized at different stages of integration into the optical modules. A total of 320 PMTs were tested and characterized, with more detailed testing performed on a small subset. Together, these results demonstrate that PMT performance is preserved through module integration and meets the acceptance criteria for deployment in P-ONE.

No PMT failures were observed. The P-ONE DAQ showed charge-resolution values comparable to those obtained with the high-resolution Picoscope. The measured charge resolution and transit-time spread meet the requirements for P-ONE's angular resolution target.

A series of high-precision measurements was performed on a subset of PMTs. Key measurements included quantum efficiency, which peaked at $27.0\pm0.7\,\%$ at 395\,nm; relative collection efficiency, which was uniform across wavelengths and ranged between 85$\%$ and 95$\%$; after-pulse rates, which were measured to be 6$\%$ with a spread of 4$\%$ across the different PMTs; and the linearity of the PMT response, which, when read out with the P-ONE electronics, extended up to about 150 photoelectrons at a gain of \num{5e6}. 

Beyond these specific results, the test setup and analysis framework developed for this campaign are now fully automated, providing a reusable tool for evaluating future photosensor candidates for P-ONE optical modules, including larger 4-inch PMTs and other sensor types. Additional detector lines are already being planned and constructed for the next iteration of the detector, the P-ONE demonstrator, building directly on the characterization methods and infrastructure established here.

\section*{AI usage statement}
Generative AI tools (large language models) were used exclusively to assist with language editing of the manuscript, including grammar and stylistic improvements. All scientific content, including figures, data analysis, results, and conclusions, was produced independently by the authors, who take full responsibility for the integrity of the work.
\acknowledgments
We thank Ocean Networks Canada for the very successful operation of the NEPTUNE observatory, as well as the support staff from our institutions without whom this experiment and P-ONE could not be operated efficiently. We acknowledge support from the Canadian Department of Fisheries and Oceans. We acknowledge the support of the Natural Sciences and Engineering Research Council of Canada (NSERC) and the Canadian Foundation for Innovation (CFI). This research was enabled in part by support provided by the BC and Prairies DRI and the Digital Research Alliance of Canada (alliancecan.ca). This research was undertaken thanks in part to funding from the Canada First Research Excellence Fund through the Arthur B. McDonald Canadian Astroparticle Physics Research Institute. P-ONE is supported by the Collaborative Research Centre 1258 (SFB1258) and the Cluster of Excellence ORIGINS, funded by the Deutsche Forschungsgemeinschaft (DFG) under Germany’s Excellence Strategy, and by the European Research Council (ERC) under the European Union’s Horizon Europe programme through an ERC Advanced Grant. We acknowledge support by the National Science Foundation. This work was supported by the Science and Technology Facilities Council, part of the UK Research and Innovation, and by the UCL Cosmoparticle Initiative. This work was supported by the Polish National Science Centre (NCN).

The authors gratefully acknowledge Markus Dittmer, Oleg Kalekin and Lasse Halve for their invaluable insights and contributions to the PMT measurements. The authors also thank the IceCube Collaboration for providing access to the testing setup used for the PMT initial characterization at RWTH Aachen.\\




\bibliographystyle{JHEP}
\bibliography{PMT_paper_test}


\end{document}